# $\alpha$-Ta Films on *c*-plane Sapphire with Enhanced Microstructure

Lena N. Majer,[1] Sander Smink,[1] Wolfgang Braun,[1] Bernhard Fenk,[1] Varun Harbola,[1] Benjamin Stuhlhofer,[1] Hongguang Wang,[1] Peter A. van Aken,[1] Jochen Mannhart,[1] Felix V.E. Hensling[1]

[1]Max Planck Institute for Solid State Research, Heisenbergstr. 1, 70569 Stuttgart, Germany

Superconducting films of $\alpha$-Ta are promising candidates for the fabrication of advanced superconducting qubits. However, $\alpha$-Ta films suffer from many growth-induced structural inadequacies that negatively affect their performance. We have therefore explored a new synthesis method for $\alpha$-Ta films which allows the growth of these films with unprecedented quality. Using this method, high quality $\alpha$-Ta films are deposited at a comparably high substrate temperature of 1150 °C. They are single-phase $\alpha$-Ta and have a single out-of-plane (110) orientation. They consist of grains $\geq 2$ μm that have one of three possible in-plane orientations. As shown by scanning transmission electron microscopy and electron energy loss studies the substrate-film interfaces are sharp with no observable intermixing. The obtained insights into the epitaxial growth of body-centered-cubic films on quasi-hexagonal substrates lay the basis for harnessing the high structural coherence of such films in various applications.

## I.     Introduction

Epitaxial interfaces between metals and oxides play a key role for a wide range of applications in electronics and catalysis. However, complex interfacial reactions, wetting, adhesion, and band-bending between the metal film and the oxide substrate often lead to difficulties in achieving epitaxial growth with low defect densities in the film, in the substrate, and at their interface.[1,2]

Driven by their potential use as superconducting layers in qubits, epitaxial films of tantalum are receiving increased attention. Tantalum crystallizes in two distinct crystallographic phases - a body-centered-cubic (bcc) $\alpha$-phase and a tetragonal $\beta$-phase. The $\alpha$-phase (lattice constant $a$ = 3.3 Å) is a superconductor with a critical temperature ($T_c$) of 4.4 K, whereas the β-phase (lattice constants $a$ = 5.3 Å , $c$ = 9.9 Å) has a $T_c$ of 0.5 K.[3] The comparatively high $T_c$ of the $\alpha$-phase is of interest for its use in quantum electronics. Indeed, there has been a breakthrough in the fabrication of long-lifetime transmon qubits made of epitaxial $\alpha$-Ta films.[4–6]

However, the epitaxial growth of $\alpha$-Ta films is demanding. The low vapor pressure of Ta implies that the deposition of Ta thin films is challenging for many film-growth techniques.[7] Therefore, Ta films are mostly grown by sputter deposition.[3–6,8–16] The resulting films, however, often suffer (1) from the formation of the unwanted $\beta$-phase, [3,6,8,9,11–15] (2) from a mixed (110) (111) orientation, [3,6,12,14,16] (3) from a high interface roughness, [4,5,12,13] and (4) from a large number of domain or grain boundaries.[6,10,13,14] Of these, (1) and (2) indirectly degrade the performance of superconducting qubits by influencing the $Ta_2O_5$ formation,[6] and (3) is the main inhibitor of a higher performance of Ta-based superconducting qubits.[4,5,17] Further, (4) has been shown to result in a decrease of the coherence length,[10] which is potentially harmful. The low loss at high frequencies of sapphire suggests it as a substrate of choice for qubit applications.[18] The growth of epitaxial Ta on $a$-plane sapphire instead of the commonly used $c$-plane sapphire has proven to be a successful strategy to achieve a pure (110) orientation. However, these films are still characterized by small domains and a correspondingly high density of domain boundaries.[6] For growth on $c$-plane sapphire, the use of a few monolayers of Nb as a nucleation layer for the Ta film growth also yielded pure (110) Ta films, and furthermore allowed a drastic reduction of the temperature necessary to suppress the formation of the $\beta$-phase. The Nb nucleation layer, however, negatively affected the crystal quality and, thus, the potential performance of respective superconducting qubit circuits.[14]

In this work we present a synthesis route providing on *c*-plane sapphire epitaxial α-Ta films, which are phase-pure, have a single out-of-plane orientation, a preferred in-plane orientation, a drastically increased grain size compared to conventional Ta films, smooth interfaces to the substrate, smooth surfaces, and are of remarkable crystal perfection. The epitaxial Ta films are grown by thermal laser epitaxy (TLE).[19] TLE offers two inherent advantages for the epitaxial growth of α-Ta: Ta is readily evaporated by the continuous wave (cw) source laser using a wavelength of ≈ 1 µm,[7,20] and high growth temperatures are readily achieved by laser heating the sapphire substrates using a wavelength of ≈ 10 µm.[7] As we will show in this work, it is these high achievable temperatures that are key to epitaxially growing α-Ta films of high quality on *c*-plane sapphire.

## II. Methods

The TLE used in this work is equipped with a Trumpf fiber laser with $\lambda = 1.07$ µm and a spot size of 1 mm², operated at an output power $P_L = 400$ W to evaporate free-standing Ta cylinders. These cylinders have a purity of 3N8, a 3 mm diameter, and a length of 6 cm. With a 60 mm distance between the hot Ta surface and the substrate, a growth rate of ≈ 0.8 Å/s is achieved under these conditions. A small melt pool at the source surface indicates a source temperature of ≥ 3000 °C. The chamber is equipped with a L-N2 shroud, such that even during deposition the background pressure is < $4 \times 10^{-10}$ hPa. The Crystek 5x5 cm² *c*-plane sapphire substrates are heated by irradiating the backside with the $\lambda \approx 10$ µm beam of a $CO_2$ laser; the substrate temperature $T_{sub}$ is measured using a backside pyrometer. Sapphire substrates are prepared as described in Ref. [21]. Films considered in this work have thicknesses between 60 nm and 120 nm as determined by x-ray reflectivity (XRR) and a Bruker DektakXT depth profilometer. Before, and after growth, the samples are characterized by in-situ reflection high-energy electron diffraction (RHEED). During growth RHEED is turned off to avoid damage to the sample.[22–24]

After growth the surface of the films is evaluated by an Asylum Cypher atomic force microscope (AFM) in tapping mode. X-ray diffraction (XRD) data were acquired with monochromated Cu-$K_{\alpha 1}$ x-rays in either a Panalytical Empyrean or a Bruker AXS D8 Discover DaVinci. Electron-backscattering diffraction (EBSD) data were acquired in Zeiss Merlin secondary electron microscope equipped with an Oxford Symmetry detector driven by Oxford AZtec software. Scanning transmission electron microscopy (STEM) specimens in the cross-sectional orientation were prepared by tripod polishing followed by argon-ion milling at liquid-N2 temperature.[25] STEM studies were conducted using a spherical aberration-corrected STEM (JEM-ARM200F, JEOL Co. Ltd.) equipped with a cold-field emission gun and a DCOR probe Cs-corrector (CEOS GmbH) operated at 200 kV. The STEM images were obtained by an ADF detector with a convergent semi-angle of 20.4 mrad and collection semi-angles of 70−300 mrad. In order to make precise measurements of lattice constants, ten serial frames were acquired with a short dwell time (2 µs/pixel), aligned, and added afterwards to improve the signal-to-noise ratio and to minimize the image distortion of high-angle annular dark-field (HAADF) images. Electron energy loss spectroscopy (EELS) acquisition was performed by a Gatan GIF Quantum ERS imaging filter equipped with a Gatan K2 Summit camera with a convergent semi-angle of 20.4 mrad and a collection semi-angle of 111 mrad. Temperature dependent resistivity was measured in van-der-Pauw geometry utilizing a Quantum Design physical property measurement system.

### III. Results

It is well established that high substrate temperatures ($T_{sub}$) are beneficial for the deposition of phase pure $\alpha$-Ta thin films.[15] Figure 1 shows the surface quality of 65 nm-thick Ta films grown at increasing $T_{sub}$. Below 600 °C amorphous growth is observed. For 600 °C ≤ $T_{sub}$ ≤ 900 °C the RHEED pattern consists of transmission spots, indicating three-dimensional (3D) growth (Fig. 1(a)). The resulting surface reveals small grains and a root mean square (RMS) value of = 2.95 nm (Fig. 1(c)). At $T_{sub}$ ≥ 1050 °C the RHEED pattern drastically changes. For growth at such high temperatures, the RHEED pattern comprises sharp surface-diffraction streaks and spots, indicating a smooth, coherent two-dimensional surface structure. Moreover, these patterns reveal a reconstruction on the epitaxially grown $\alpha$-Ta film. The spots span six times the fundamental spacing, with 5 regular reconstruction streaks in between each one, thereby revealing a six-fold superstructure.

The resulting surface topography of the film grown at 1050 °C is comparatively smooth with an RMS = 1.42 nm (Fig. 1(d)). The grains are about ten times wider than the film thickness and their lateral dimension is by far larger than that of state-of-the-art films.[4,14] The elongated grains are aligned along three in-plane orientations which differ by 120 ° (compare Fig.4).

Upon increasing $T_{sub}$ further to 1150 °C the grains grow even larger to attain µm sizes and interconnect (Fig. 1(e)), resulting in a reduced surface roughness with RMS = 0.52 nm.

For $T_{sub}$ = 1250 °C, the grains are even larger, but they get separated by deep trenches. Interestingly, on top of these individual grains, steps of ≈ 2.3 Å are visible in the AFM micrographs (Fig. 1(f)), this step height corresponds to the $d_{110}$ spacing of α-Ta. The surface morphologies of films grown at $T_{sub}$ > 1250 °C become more irregular (Sup. Fig. 1).

Having found that $T_{sub}$ = 1150 °C results in smooth and closed films of α-Ta, we now turn to the question whether these films also have favorable microstructural properties. The XRD $\theta$-$2\theta$ scans taken on such films (Fig. 2(a)) reveal a (111) orientation for the film grown at $T_{sub}$ = 900 °C. At $T_{sub}$ = 1050 °C, additionally, the expected 110 and 220 peaks of epitaxial α-Ta are observed. For $T_{sub}$ = 1150 °C and 1250 °C only 110 and 220 peaks of epitaxial α-Ta are observed. These measurements thus uncover that for the growth conditions used, growth at $T_{sub}$ > 1050 °C is required to grow α-Ta in single out-of-plane orientation on c-plane sapphire. Despite the mixed orientation of the films grown at $T_{sub}$ = 1050 °C, all films were found to exhibit distinct Laue oscillations, which indicate a sharp substrate-film interface, as well as a good crystallinity of the film, and a homogenous film thickness (Fig. 2(b)).[26] Figure 2(b) further reveals a slight shift of the 110 peak to lower angles with increasing $T_{sub}$. In Table 1 the resulting lattice parameters are compared to the bulk value.[27] The slight increase in the lattice parameter is likely the result of the increased formation of Ta vacancies at increased $T_{sub}$, which are frozen by the rapid cooldown.

Fig. 2(c) shows the rocking curves measured at the 110 peak for all films exhibiting this orientation. Astoundingly, the rocking curve peak is very sharp for all three $T_{sub}$. Indeed, its full-width half-maximum (FWHM) values are $0.038° \leq \Delta\omega \leq 0.041°$. The rocking curve peaks are, thus, more than two times narrower than those of the best Ta-films published, which actually were obtained on a-plane sapphire.[28] Interestingly, grain size and single orientation seem to not significantly impact the perfection of the crystal quantified by the FWHM.

As expected from the distinct Laue oscillations XRR reveals extensive Kiessig fringes for the films grown at $T_{sub} \geq 1050$ °C. These are a testament to the sharp substrate-film interface and the high quality of the film surface.

To gain insight into the epitaxial relationship between the film and the substrate, and to decipher the in-plane orientation of the films, we present in Fig. 3 the XRD pole figures of a film grown at $T_{sub}$ = 1150 °C and a substrate. The pole figures clearly reveal a Pitsch-Schrader epitaxial relation[29] with very sharp reflections. Patterns reflecting other orientations[16] are completely absent. For the Pitsch-Schrader orientation of (110) Ta on (0001) Al$_2$O$_3$, the in-plane direction [001] of the bcc Ta is parallel to the sapphire direction [1$\bar{1}$00] or its equivalents [01$\bar{1}$0] and [$\bar{1}$010], yielding three degenerate in-plane orientations. The corresponding arrangements of the atomic lattices are schematically depicted in Fig. 4. The Ta in-plane vectors [1$\bar{1}l$] and [$\bar{1}$1$l$] ($l$ = 1 or $\bar{1}$) are at an angle of 5.26° to the [$\bar{1}$010] direction or its equivalents of the hcp $c$-plane sapphire substrate. Interestingly, as shown in Fig. 3(a) for $T_{sub}$ = 1150 °C, the diffraction spots at $\phi$ = 0 ° and $\phi$ = 180 ° are more pronounced than the others, disclosing one orientation is preferred over the other two. The films grown at 1050 °C and 1250 °C have no strongly preferred orientation (Sup. Fig. 2). This can be quantified by the relative intensities of the underlying $\phi$-scans found in Table 2. The peaks are labeled as in Sup. Fig. 3. The degeneracy of the in-plane orientations is thus lifted in samples grown at $T_{sub}$ = 1150 °C, which again shows that the temperature window for the ideal growth of epitaxial $\alpha$-Ta is narrow. EBSD data (Sup. Fig. 4) show that the domains with preferred orientation merge to form very large single-crystalline domains. This may provide a starting point for realizing single-crystalline bcc films on hexagonal or triangular substrates.

Direct atomic-scale information on the crystal lattices and the substrate-film interfaces has been obtained from extensive STEM studies of the films. Figure 5 (a) shows a wide-field HAADF image of a cross-section through the film and substrate. The micrograph reveals a uniform thickness. The image also shows faint white lines which likely indicate grain boundaries that meet at a 120° angle. The atomic resolution images shown in Figs. 5(c) and (d) directly confirm the high crystal quality on the nanoscale as well as the atomically sharp interface between substrate and film. The rotation angle between the two images is ≈ 5.2°, providing direct real-space evidence for the film-substrate atomic arrangement depicted in Fig. 4. For clarity the atomic arrangement of Fig. 5(c) is additionally depicted in Fig. 5(e). The atomic sharpness of the interface is also revealed by the EELS maps in Fig. 5 (b), in which no intermixing between film and substrate ions can be discerned, this despite the high $T_{sub}$ at which the films were grown.

Figure 6 shows the resistivity of the $\alpha$-Ta films grown at different $T_{sub}$ as a function of temperature. For comparison, the figure also shows the behavior of a bulk $\alpha$-Ta sample as published in Ref. [30]. Films grown at $T_{sub} \geq 1050$ °C show $T_c$ close to the bulk value. In fact, the $T_c$ of the samples grown at 1050 °C and 1150 °C is the same as the bulk value ($\approx 4.4$ K, see inset of Fig. 6). It is to be noted that the sample grown at 1050 °C is by far the thickest sample (120 nm). Of all samples it has the most noteworthy residual resistivity ratio (RRR) $\rho_{300 K}/\rho_{5 K}$ of 33, which is much higher than any literature values for thin films. At the identified ideal growth temperature (1150 °C), the RRR $\rho_{300 K}/\rho_{5 K}$ of the 80 nm thick film is 13. This value is comparable to literature values.[6] In our case, however, this value is exhibited by films that are an order of magnitude thinner than the reference films, which greatly increases the relative contribution of the electron scattering at the film surface and at the film-substrate interface to the sample resistance at low temperatures. The RRR $\rho_{300 K}/\rho_{5 K}$ of the film grown at 1250 °C is markedly worse (4.75) and its $T_c$ is lowered. The observed grain separation at this temperature, thus, has a negative influence on the superconducting properties. The film grown at 900 °C, which is 111 oriented, has a much higher resistivity, a low RRR $\rho_{300 K}/\rho_{5 K}$ (1.6) and its $T_c$ is shifted to the limit of the measurement setup. This underlines the importance of the ability to grow 100 oriented $\alpha$-Ta films at $T_{sub} \geq 1050$ °C. The partial 111 orientation at $T_{sub} = 1050$ °C, however, has no significant negative effect on the superconducting properties. Contrary, the RRR for this film is remarkably high.

## IV. Summary & Conclusion

This aim of this work is to find a new route to the epitaxial growth of $\alpha$-Ta thin films, to avoid growth-related deficits some of which negatively affect the performance of superconducting devices. These defects include unwanted $\beta$-phase formation, a mixed (110) (111) orientation, a high interface roughness, and large numbers of grain boundaries. We have found that TLE offers the necessary parameter range for the growth of defect-poor $\alpha$-Ta films, as Ta is readily evaporated by laser heating and high substrate temperatures are achievable by $CO_2$-laser heating while maintaining low background gas pressures.[7] Our studies establish that substrate temperatures $T_{sub} > 1050$ °C are required to ensure a single (110) orientation of $\alpha$-Ta films. The preferred substrate temperature for the growth of Ta films that have large domains, an atomically sharp interface with the underlying substate, a singular out-of-plane and a preferred in-plane orientation is $T_{sub}$ = 1150 °C. These films are closed, smooth, and free of $β$-Ta impurities.

## Acknowledgements

The authors would like to thank Hans Boschker for scientific discussions and Gunther Richter, and Gerd Maier for experimental support. The authors acknowledge Ute Salzberger for assistance with TEM sample preparation.

## Conflict of Interests

The authors have no conflict of interest to declare

## Data availability

The data supporting the finding of this study are available within the paper, its supplementary, and upon reasonable request.

**Table 1** Comparison of the bulk lattice parameter ($a$)[27] and the lattice parameter of the films grown and different substrate temperatures obtained from Fig. 2 (b).

|   | bulk | $T_{sub}$ = 1050 °C | $T_{sub}$ = 1150 °C | $T_{sub}$ = 1250 °C |
|---|---|---|---|---|
| $a$ | 3.303 Å | 3.304 Å | 3.306 Å | 3.311 Å |

**Table 2** Relative peak intensities extracted from the respective ϕ-scans labeled as in Sup. Fig. 3 and calculated from the sum of the respective peaks.

| $T_{sub}$ | 1050 °C | 1150 °C | 1250 °C |
|---|---|---|---|
| A (002) | 1.00 | 0.29 | 0.81 |
| B (002) | 0.92 | 1.00 | 0.50 |
| C (002) | 0.95 | 0.52 | 1.00 |
| A (110) | 0.95 | 0.31 | 0.78 |

| B (110) | 1.00 | 1.00 | 0.45 |
| C (110) | 0.96 | 0.53 | 1.00 |

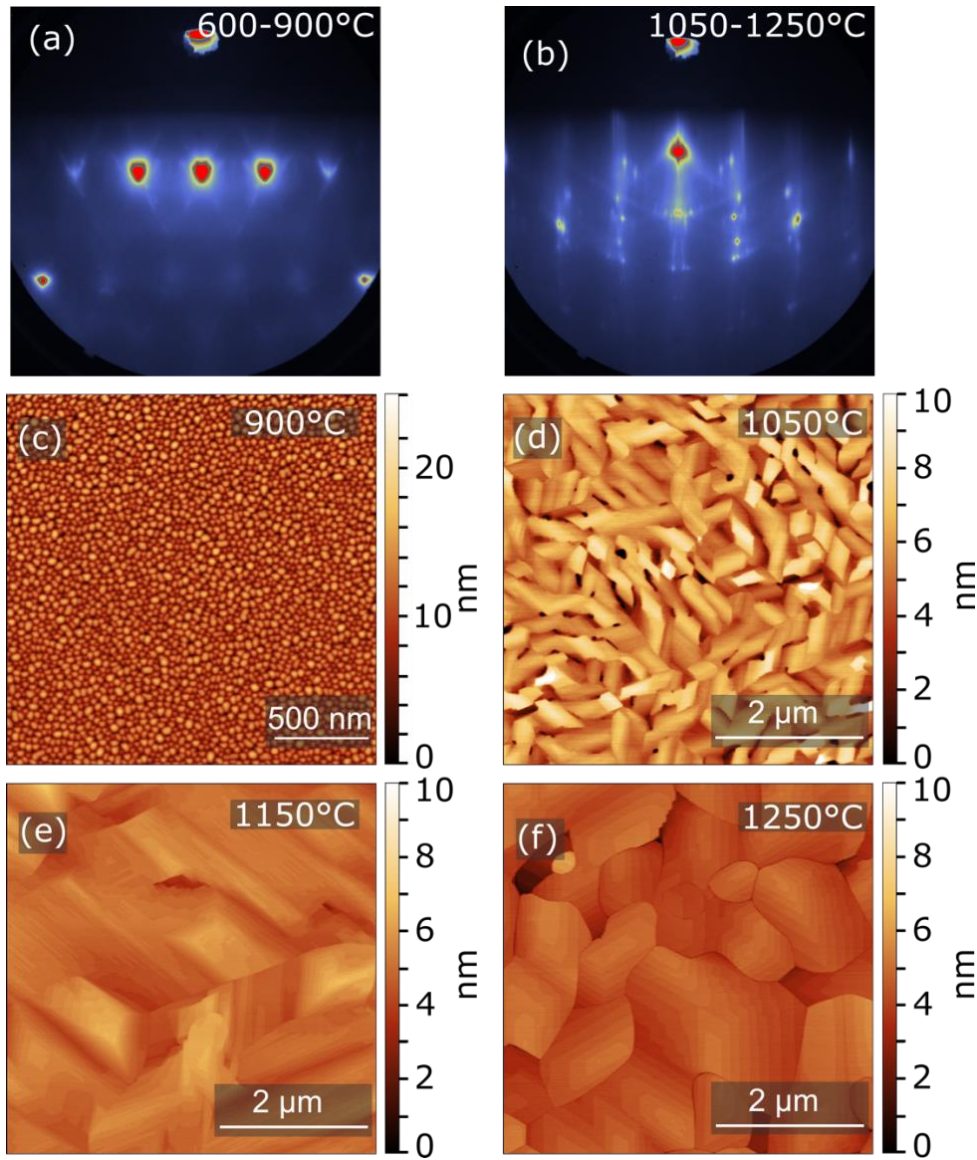

**Figure 1** (a) RHEED image representative of α-Ta film grown between 600 °C and 900 °C revealing a 3D surface structure. (b) RHEED image representative films grown at 1050 °C ≤ $T_{sub}$ ≤ 1250 °C, revealing a drastically improved surface structure. (c) – (f) AFM micrographs showing the surface topography of films grown at 900 °C, 1050 °C, 1150 °C, and 1250 °C, respectively.

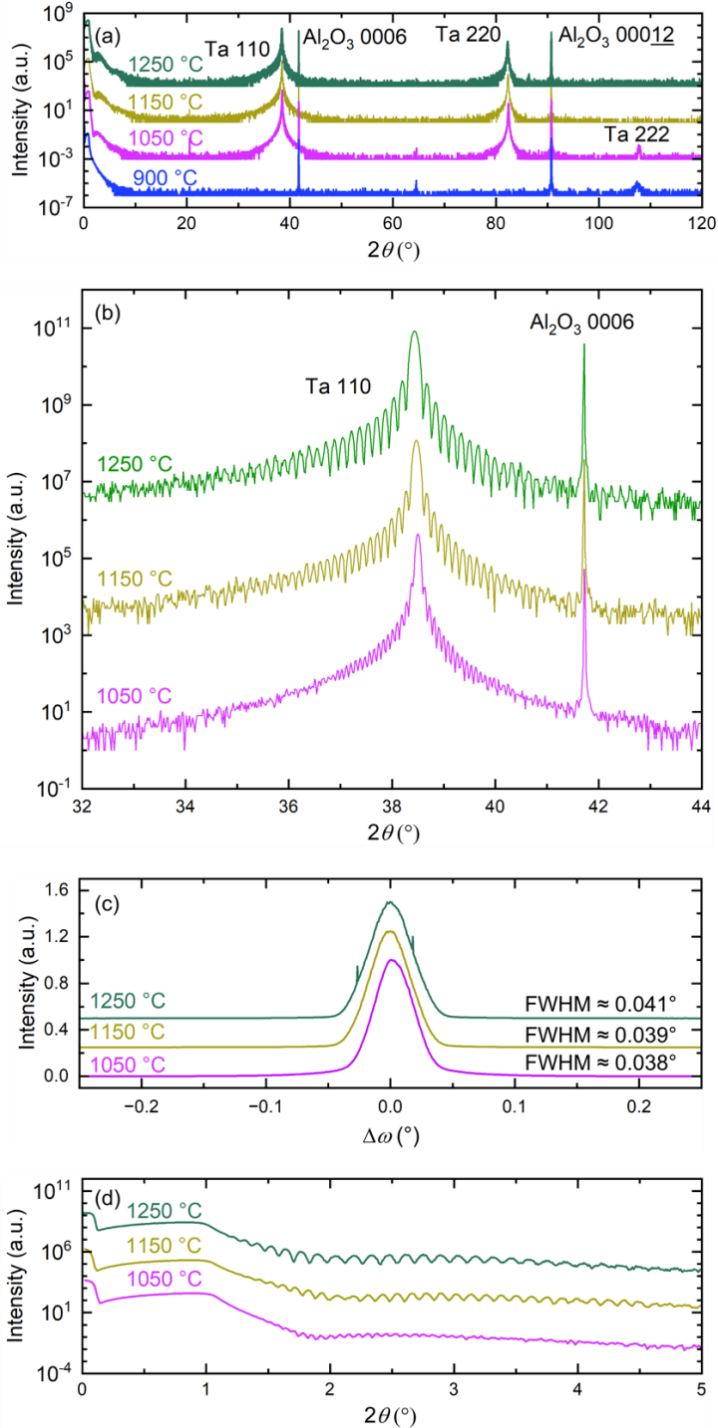

**Figure 2** (a) Wide range θ-2θ scan (measured by Cu-K$_{\alpha 1}$ x-rays) of α-Ta films grown at 900 °C, 1050 °C, 1150 °C, and 1250 °C, showing the expected Ta 110 and 220 peak in addition to the c-plane sapphire substrate peaks (sometimes including the commonly observed forbidden sapphire peaks).[31,32] The film grown at 1050 °C is of mixed orientation as evidenced by the notable 222 peak. The film grown at 900 °C is of pure 111 orientation. (b) Zoom in on the α-Ta 110 peak of (a) revealing Laue-oscillations. (c) Rocking curve around the 110 peak and (d) XRR of the same films.

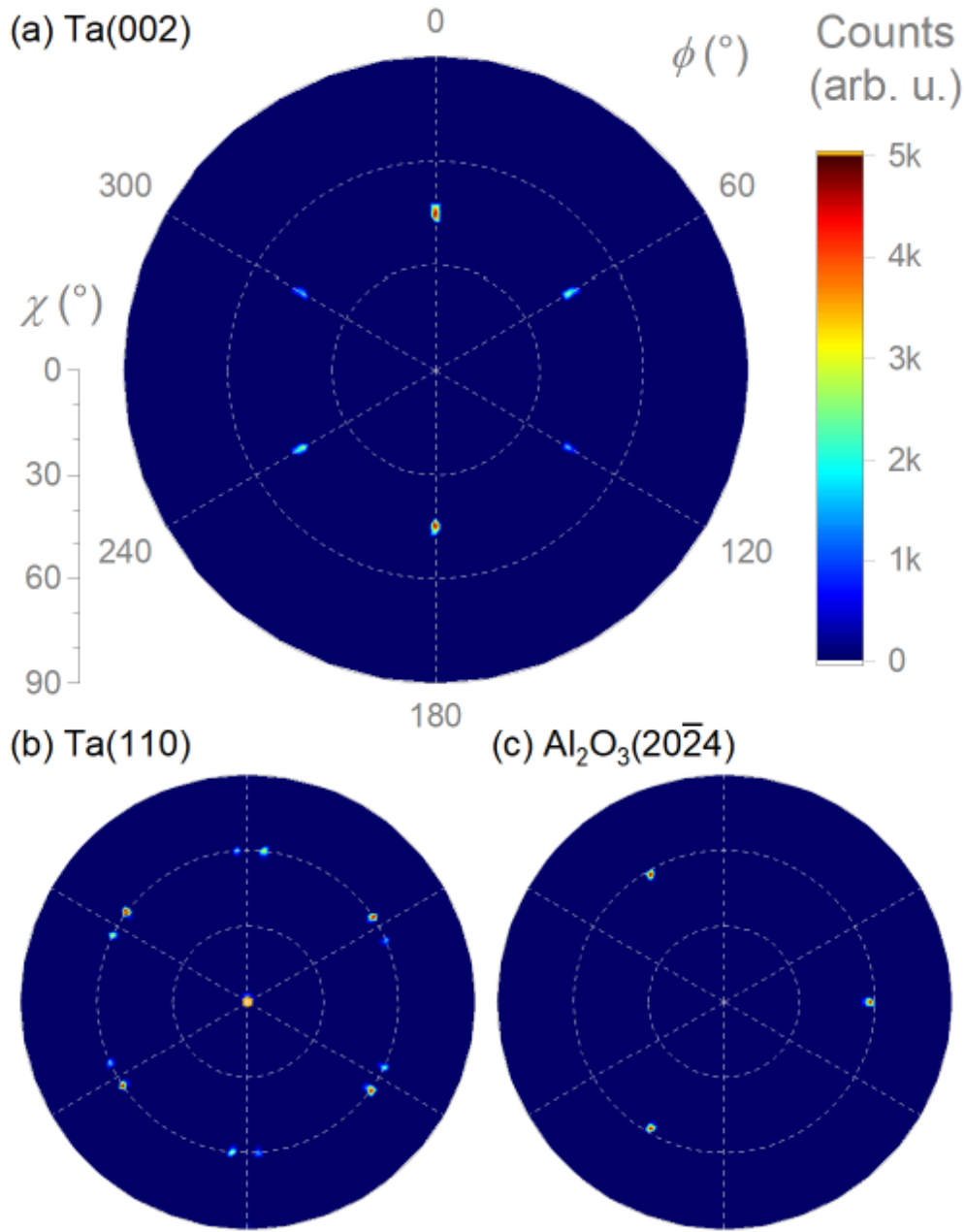

**Figure 3** XRD pole figures of (a) α-Ta along (002) () at 2θ = 55.6°, (b) α-Ta along (110) at 2θ = 38.5°, and (c) sapphire along (20$\bar{2}$4) at 2θ = 52.6°. The azimuthal angle φ corresponds directly to the respective angle in the polar plot, and the polar angle χ is given as the radial distance according to the scale in (a). The linear intensity scale is color-coded.

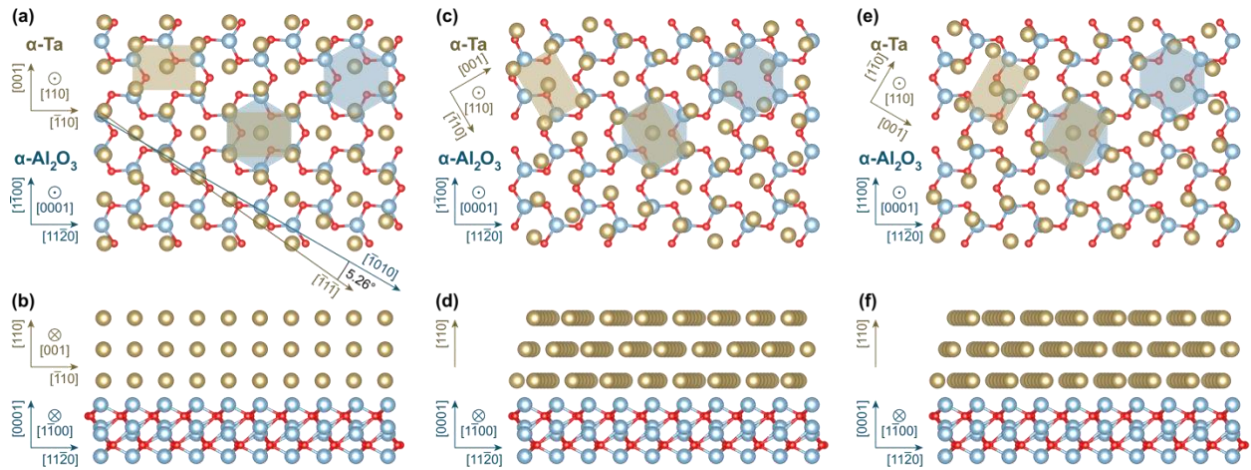

**Figure 4** Illustrations of the atomic arrangements of α-Ta films on c-plane sapphire as viewed from the top (top row) and side (bottom row), for the three possible epitaxial relationships.

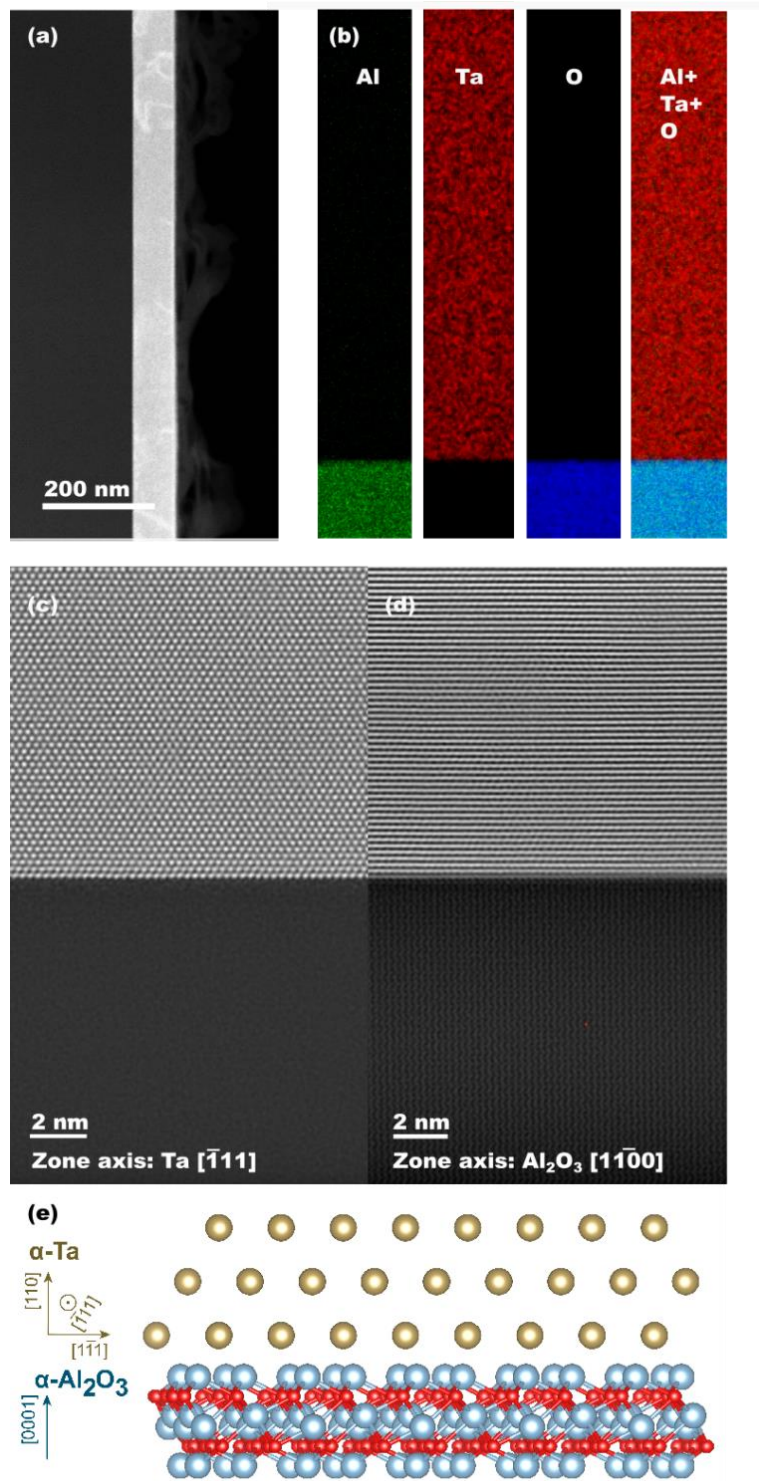

**Figure 5** (a) Wide field-of-view high angle annular dark field (HAADF) image of an α-Ta film grown at T = 1150 °C on c-plane sapphire. The growth direction is from left to right. (b) EELS data with growth direction from bottom to top. Atomic resolution HAADF along (c) the [$\bar{1}11$] zone axis of Ta and (d) the [$1\bar{1}00$] zone axis of sapphire. The angle between the zone-axes used for (c) and (d) is ≈ 5.2 °. (e) Illustration of the atomic arrangement observed in (c).

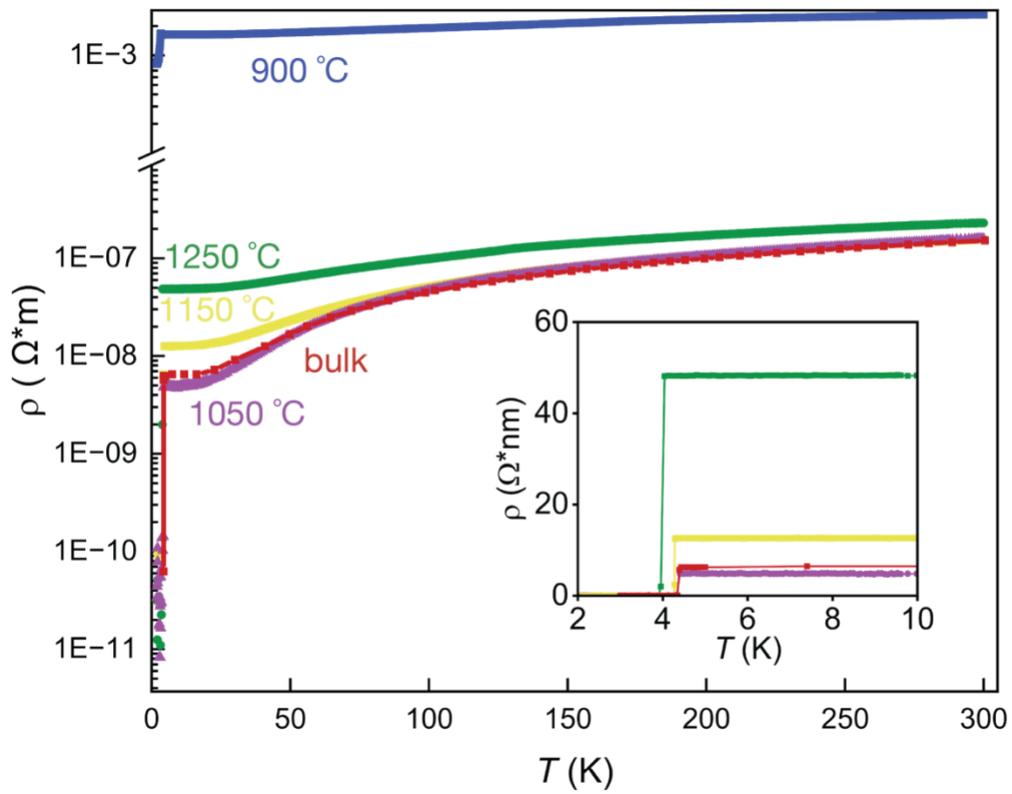

**Figure 6** Resistivities of α-Ta films grown at various substrate temperatures and of a bulk sample (taken from [30]) measured as a function of temperature.